\documentclass[10pt,twocolumn,twoside,lineno]{revtex4-2}

\usepackage{color}
\usepackage{graphicx}
\usepackage{dcolumn}
\usepackage{bm}
\usepackage{array}
\usepackage{amsmath}

\begin{document}

	
\title{Morphogenesis of sound creates acoustic rainbows}

\author{Rasmus E. Christiansen}
\author{Ole Sigmund}

\affiliation{Department of Mechanical Engineering, Technical University of Denmark, Nils Koppels Allé, Building 404, 2800 Kongens Lyngby, Denmark}

\author{Efren Fernandez-Grande} 

\affiliation{Department of Electrical Engineering, Technical University of Denmark, Ørsteds Plads, Building 352, 2800 Kongens Lyngby, Denmark}




\begin{abstract}
Sound is an essential sensing element for many organisms in nature, and multiple species have evolved organic structures that create complex acoustic scattering and dispersion phenomena to emit and perceive sound unambiguously. To date, it has not proven possible to design artificial scattering structures that rival the performance of those found in organic structures. Contrarily, most sound manipulation relies on active transduction in fluid media rather than relying on passive scattering principles, as are often found in nature. In this work, we utilize computational morphogenesis to synthesize complex energy-efficient wavelength-sized single-material scattering structures that passively decompose radiated sound into its spatio-spectral components.  Specifically, we tailor an acoustic rainbow structure with "above unity" efficiency and an acoustic wavelength-splitter. Our work paves the way for a new frontier in sound-field engineering, with potential applications in transduction, bionics, energy harvesting, communications and sensing.
\end{abstract}



\maketitle

\section*{Introduction}
Rainbows, the spatial decomposition of white light into its spectral components, are frequently observed when light propagates through dispersive media – prisms, droplets, or similar \cite{BOOK_OPTIKS}. The analogous phenomenon of spatio-spectral decomposition of sound in free-space, where waves oscillating at different frequencies propagate in different directions, is less well known. The decomposition of sound into spectral components can occur in confined media \cite{ZHU_ET_AL_2013,Smolyaninova_et_al_2010}, where arrays of resonant structures “trap” sound at different positions in space depeding on frequency \cite{Karlos_et_al_2020}, as in waveguides or reactive silencers \cite{BOOK_MUFFLERS}, in solid and/or fluid mixtures \cite{White_et_al_2005}, or in acoustic circulators utilizing non-reciprocity \cite{Fleury_et_al_2014}. Acoustic metamaterials have shown remarkable potential for manipulating sound fields \cite{Xie_et_al_2016,Liang_et_al_2012}, using arrays of acoustical \cite{Hakansson_2006, Bozhko_et_al_2017} or acousto-mechanical resonant structures \cite{Esfahlani_et_al_2016} as also exploited by the so-called leaky wave antennas \cite{Naify_2013} or using topological structures \cite{Chaplain_et_al_2020, Tian_2020}, which lead to unusual propagation phenomena, including strong acoustic dispersion.

Nonetheless, the occurrence of spectro-spatial decomposition of sound (not requiring unusual acoustic dispersion or exotic propagation phenomena) can be observed in nature, where living organisms utilize sound to relate to their environment. Examples include the mammalian pinnae \cite{BOOK_HEARING,Hofman_et_al_1998}, noise emitting bats \cite{Linnenschmidt_et_al_2016}, cetacea \cite{Branstetter_et_al_2012,Lammers_et_al_2003,Miller_2002} and others \cite{Dantzker_1999}. The outer ear of primates is a close example to our species \cite{Coleman_2004}: its intricate geometry produces complex interference phenomena, which result in unique spectro-spatial cues that aid the localization of sound sources. However, to date, ad-hoc man-made structures have failed in achieving a controlled spectro-spatial decomposition of sound based solely on scattering. It is this gap of knowledge that motivates our efforts. 

Our work focuses on a general unifying framework that enables the design of structures that fully control the spectro-spatial decomposition of emmitted/received sound. We present a single-material passive acoustic-scattering structure, systematically tailored using computational morphogenesis to convert a broadband white-noise signal emitted by a single monopolar source into an acoustic rainbow with "above unity" radiation efficiency across the entire targeted frequency band by simultaneously shaping the emission pattern and maximizing the emission efficiency (Fig. 1, 2 and 4). Subsequently, we use the same approach to design a lambda-splitter capable of dividing a broadband signal into two well-defined spatio-spectrally separated frequency bands (Fig. 3). Both structures are based solely on the scattering of sound from hard surfaces and work across 50\% frequency bands under free field conditions. Through experiments we show close agreement between the numerical predictions and real world operation, demonstrating the strength and accuracy of the design method. The Supplementary Information includes videos of simulations illustrating the workings of the two devices, showing the near field emission patterns as a function of frequency and illustrating the acoustic rainbow as heard by an observer at different observation angles. Due to the reciprocity of the physical system, the properties of emission and reception may be trivially reversed. Our work thus paves the way for a new frontier in tailoring the emission and reception of sound fields. The results presented in the following are also of relevance to other disciplines concerned with the sensing and emission of wave fields. 

\begin{figure*}
	\centering
	\includegraphics[width=0.99\linewidth]{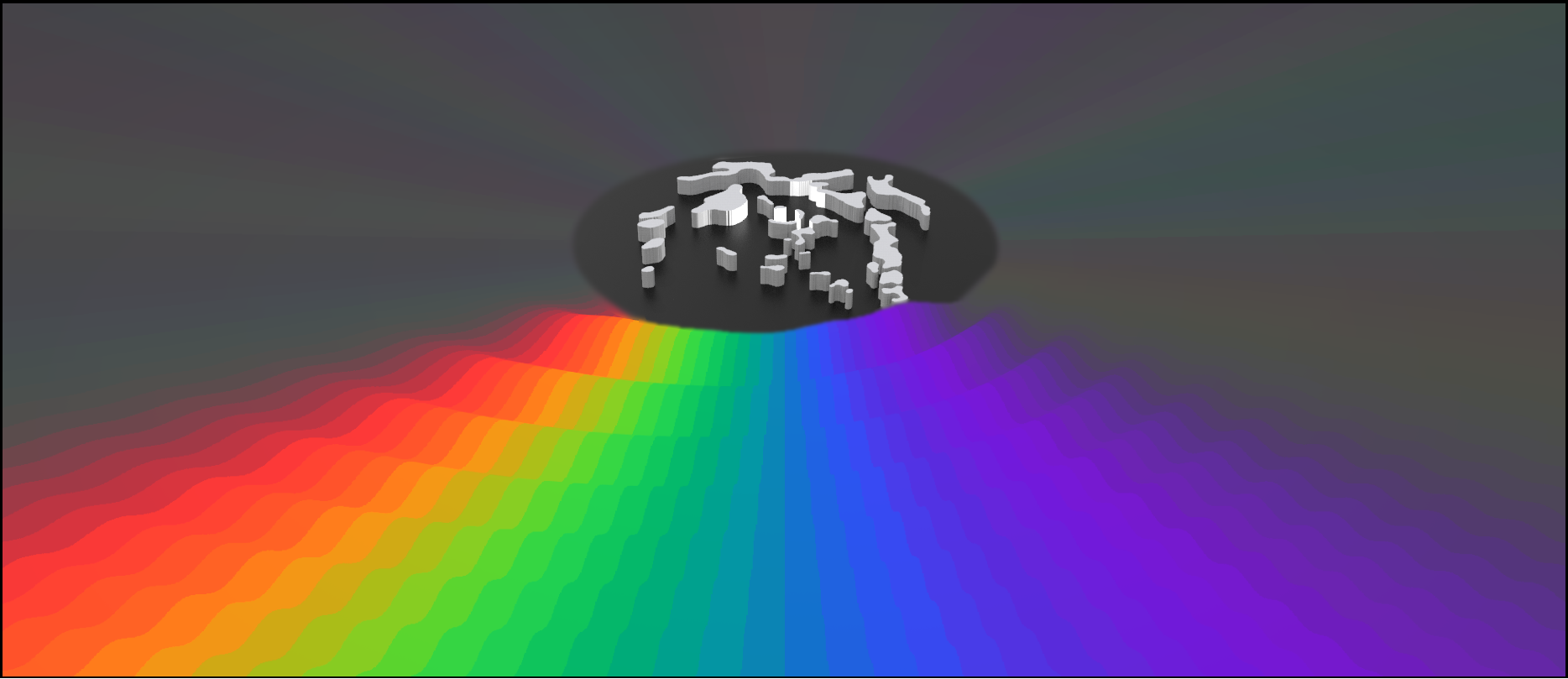}
	\caption{Visualization of the Acoustic Rainbow Emitter (ARE). The morphogenetical TopOpt method shapes the scattering inclusions that compose the, shown as gray material. When the ARE is excited by mono-polar source emitting broad-band white noise, the radiated sound creates an acoustic rainbow. The source is positioned at the center of the emitter (illustrated using white light), and driven with equal power at all frequencies from 7600 Hz to 12800 Hz. In the figure, the experimentally measured acoustic output (far field) is mapped to the visible spectrum of light by its magnitude and frequency content in the full 360$^{\text{o}}$ surrounding the ARE. (See the Supplementary Information for a detailed description).}
	\label{FIG:ACOUSTIC_RAINBOW_ILLUSTRATION}
\end{figure*}

Inspired by the natural occurrences of passive “sound-shaping” devices, we modify our earlier morphogenetic design framework \cite{Christiansen_Grande_2016} for tailoring passive acoustic scattering structures with dimensions on the order of a few wavelengths, to also control the near- and far-field radiation pattern depending on the frequency content of the sound field. The modified framework is set up to design scattering structures, which exert precise spatio-spectral control of the radiation and simultaneously improve the impedance matching between the source and surrounding medium, acting as an acoustic antenna, leading to "above unity" radiation efficiency relative to the free-space emission of the source. 

In short, our design approach consists of the iterative process of redistributing acoustically reflecting material in an air background inside a specified region of space, $\Omega_D$, in order to match the sound-field emitted from the device under design $\vert p(\textbf{x}) \vert$ to a desired target emission pattern $\vert p_{\mathrm{target}}(\textbf{x}) \vert$ across a specified frequency band elsewhere in space $\Omega_t$. This goal may in short be written as the figure of merit 

\begin{equation} \label{EQN:FOM}
\Phi = \int_{\Omega_t} \left( \vert p(\textbf{x}) \vert^2 - \vert p_{\mathrm{target}}(\textbf{x}) \vert^2 \right)^2 \mathrm{d}\textbf{x},
\end{equation}

Using this figure of merit and the corresponding adjoint sensitivity analysis  to obtain gradient information \cite{TORTORELLI_ET_AL_1994}, combined with an iterative design update scheme exploiting a gradient-based optimization algorithm (MMA \cite{Svanberg_2002}), devices can be efficiently tailored to emit sound in a wide range of unusual patterns and magnitudes by appropriate selection of $p_{\mathrm{target}}(\textbf{x})$. In this work we consider a two-dimensional spatial setting for simplicity, however, neither the numerical design approach nor the practical realization of the designed devices hinders direct extension to three-dimensions. The prescribed maximum size of the design dictates the minimum frequency it can effectively operate at when relying on scattering. We considered structures with a size of 10 cm x 10 cm and smaller (similar to the mammalian pinnae), which can operate effectively from $\approx4$ kHz and upwards. Given dispersion free materials a given design can be trivially rescaled to shift the operational bandwidth. A more detailed description of the design procedure, the numerical methods used for simulations, the experimental setup, animations and audio examples are provided in the Supplementary Information and references therein. 

Our work takes advantage of the rapid growth in computational power, which in recent years have enabled large-scale modelling and synthesis of sound as a full-wave phenomenon in complex domains and geometries. Further, advances in modelling and production techniques have enabled the use of morphogenetic methods such as TopOpt \cite{BOOK_TOPOPT_BENDSOE} in the design of practically realizable structures, ranging in size from decameters to nanometers \cite{Aage_et_al_2017,Christiansen_Grande_2016,Alexandersen_et_al_2016,Jensen_2011,Duhring_2008,Wadbro_Berggren_2006}. Topology optimization has previously been applied in acoustics to tailor phononic crystals and metamaterials \cite{Li_et_al_2019}, designing for effective negative material parameters \cite{Dong_et_al_2019, Dong_et_al_2020}, tailoring hyperbolic metamaterials \cite{Rong_et_al_2019} and topological insulators \cite{Christiansen_et_al_2019,Chen_et_al_2021} as well as for acoustic-structure interaction problems \cite{Dilgen_et_al_2019} for filtering \cite{DILGEN_2024} and sound-insulation \cite{COOL_2023} among others. Topology optimization, combined with accurate modelling of the wave-nature of sound \cite{BOOK_FUNDAMENTAL_ACOUSTICS} and the availability of advanced production techniques like 3D prinitng provde near-unlimited design freedom, enabling the design of novel materials and non-intuitive structures hitherto deemed unachievable.

\section*{Results}

\subsection*{The Acoustic Rainbow Emitter}

Figure~\ref{FIG:ACOUSTIC_RAINBOW_ILLUSTRATION} presents the experimentally measured sound power emitted from a single centrally-placed source and shaped by the acoustic rainbow emitter (ARE) designed using the approach detailed in the Supplementary Information by minimizing the figure of merit in eq.~\eqref{EQN:FOM} for 20 equidistant frequencies across the interval 7.6 kHz to 12.8 kHz. The illustration of the measured sound power is made by mapping the sound-field according to its angle-dependent frequency content to the visual spectrum of light. The ARE is shown to scale relative to the wavelengths of the sound field. 

\begin{figure*}
	\centering
	\includegraphics[width=.6\linewidth]{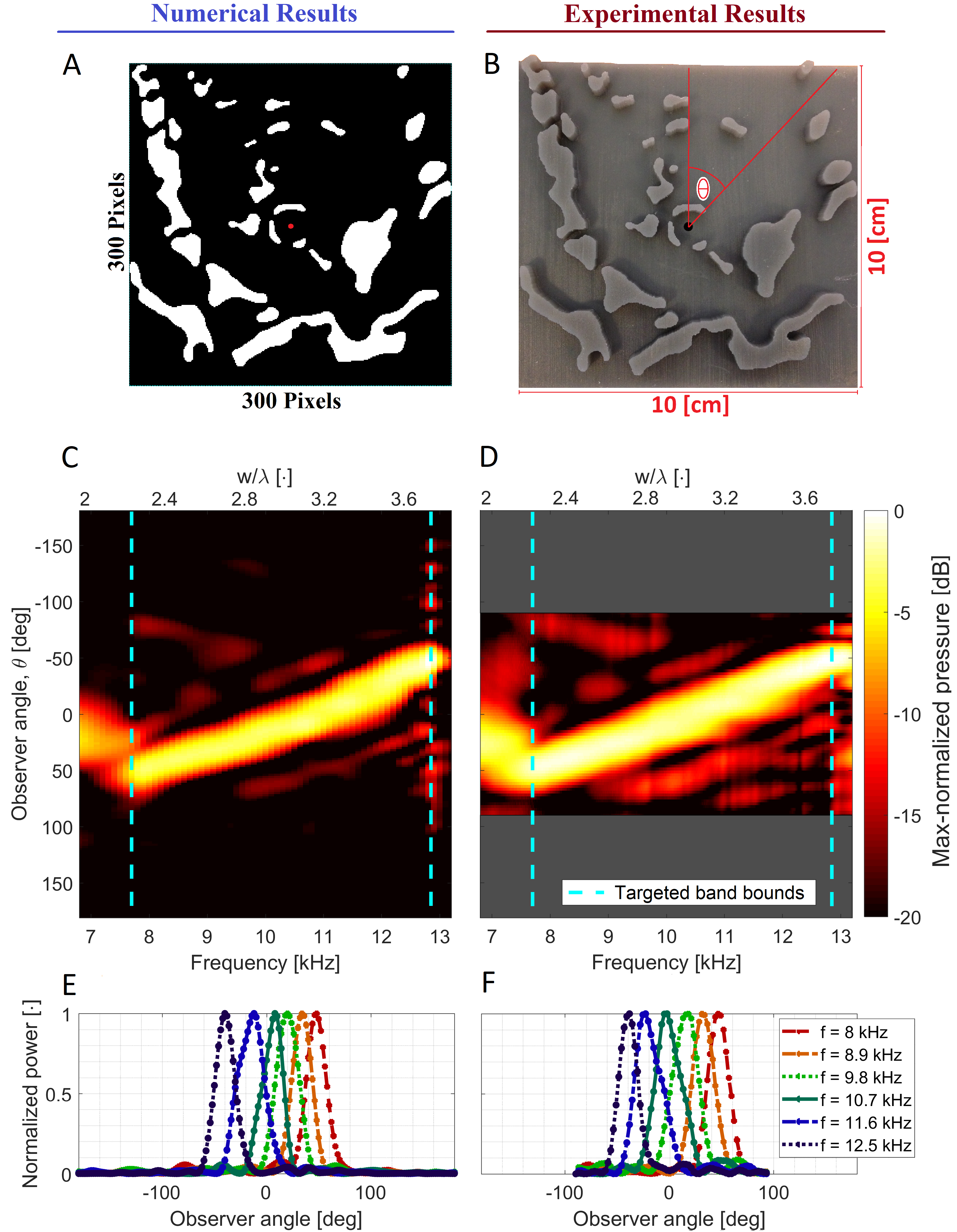}
	\caption{ARE geometry and emitted sound pressure: (left column) numerical results and (right column) experimental results. (A) Blueprint for the ARE designed by our morphogenetic computational approach, the mono-polar source position is indicated by a red dot. (B) Experimental construction of the ARE (3D printed design), sample size and angular convention are overlaid. (C-D) Maps of the max-normalized far field pressure. (C) Simulated emission, the spatio-spectral decomposition is observed as different frequencies are emitted in different angles. (D) Experimental measurements of the sound pressure emitted by the ARE, showing an almost perfect agreement with simulations (see Supplementary Information for the data treatment procedure). (E-F) Per frequency max-normalized far field power. (E) Simulated sound emission. (F) Experimental measurements performed on the 3D printed sample.}
	\label{FIG:ACOUSTIC_RAINBOW_INVESTIGATION}
\end{figure*}

The device blueprint for the ARE is presented in Fig.~\ref{FIG:ACOUSTIC_RAINBOW_INVESTIGATION}a with solid(void) regions shown using white(black) and the monopolar source position indicated by the red dot. The blueprint consists of numerous intricately shaped sound-hard features among which are found three centrally placed reflectors near the source, guiding its emission throut three channels of different width. A reflector-like ridge with six narrow channels through it is situated behind and to the left of the source. This ridge serves to reflect and guide the sound waves in the specified directions, as the channels create acoustic paths of different length, which dictate how the waves interfere depending on their frequency in order to progressively guide waves of lower frequency towards the right (+$\theta$), and of higher frequency towards the left (-$\theta$). Due to complex interplay between the many design features, it is difficult (if not impossible) to fully isolate the effect of any individual feature. Yet, it is later shown that removing any one individual feature changes the radiation pattern significantly, demonstrating that the device operates through a complex interplay of scattering and interference between the different features. Figure~\ref{FIG:ACOUSTIC_RAINBOW_INVESTIGATION}b presents an image of the 3D-printed experimental specimen with the physcial dimensions and the adopted angular convention overlaid. Figure~\ref{FIG:ACOUSTIC_RAINBOW_INVESTIGATION}c presents the numerically simulated radiated sound-pressure magnitude clearly revealing the rainbow emission pattern, with the angle of maximum emitted acoustic power varies continuously from -50$^{\text{o}}$ to 50$^{\text{o}}$ as the frequency shifts from 7.6 kHz to 12.8 kHz. The vertical dashed lines specify the bounds of the frequency-band targeted in the design process. From the data it is clear that the spatial separation of the spectral components is successful, achieving differences in acoustic power of at least one order of magnitude between the main emission lobe and any side lobes. Comparing the simulated data to the experimental measurements shown in Fig.~\ref{FIG:ACOUSTIC_RAINBOW_INVESTIGATION}d clear agreement is observed. In both the simulated and measured data weak side-lobes are observed, these are caused by a fraction of the emitted acoustic energy not being effectively scattered in the prescribed direction by the optimized design. The normalized far-field power as a function of emission angle, for six equidistant frequencies, is presented in Fig.~\ref{FIG:ACOUSTIC_RAINBOW_INVESTIGATION}e (simulation) and Fig.~\ref{FIG:ACOUSTIC_RAINBOW_INVESTIGATION}f (experiment), respectively. The regular angular shift in the emission direction as a function of frequency is clearly observed, as is a substantial difference in main- to side-lobe power of approximately 20 dB. In addition to achieving the desired spatio-spectral response, the ARE achieves "above unit" radiation efficiency across the entire frequency band of operation in terms of the total sound power emitted by the source relative to the source radiating into free space (see the section on emission efficiency and Figure~S2 in the Supplementary information for details). The high radiation efficiency is obtained by the morphological design method as it implicitly matches the impedance of the source to the surrounding medium through the scattering structures constituting the ARE in order to match the prescribed target field. Such "above unity" radiation efficiency may be contrasted with previous work on the design of an acoustic prism using an array of acousto-mechanical resonant structures \cite{Esfahlani_et_al_2016}, where a radiation efficiency of a few percent was reported. Our results thus prove that the proposed design framework enables achieving highly efficient spatio-spectral decomposition of sound through passive acoustic scattering from systematically engineered single-material acoutically-hard surfaces, as the emission/reception properties are tailored through a target figure of merit.

\begin{figure*}
	\centering
	\includegraphics[width=.6\linewidth]{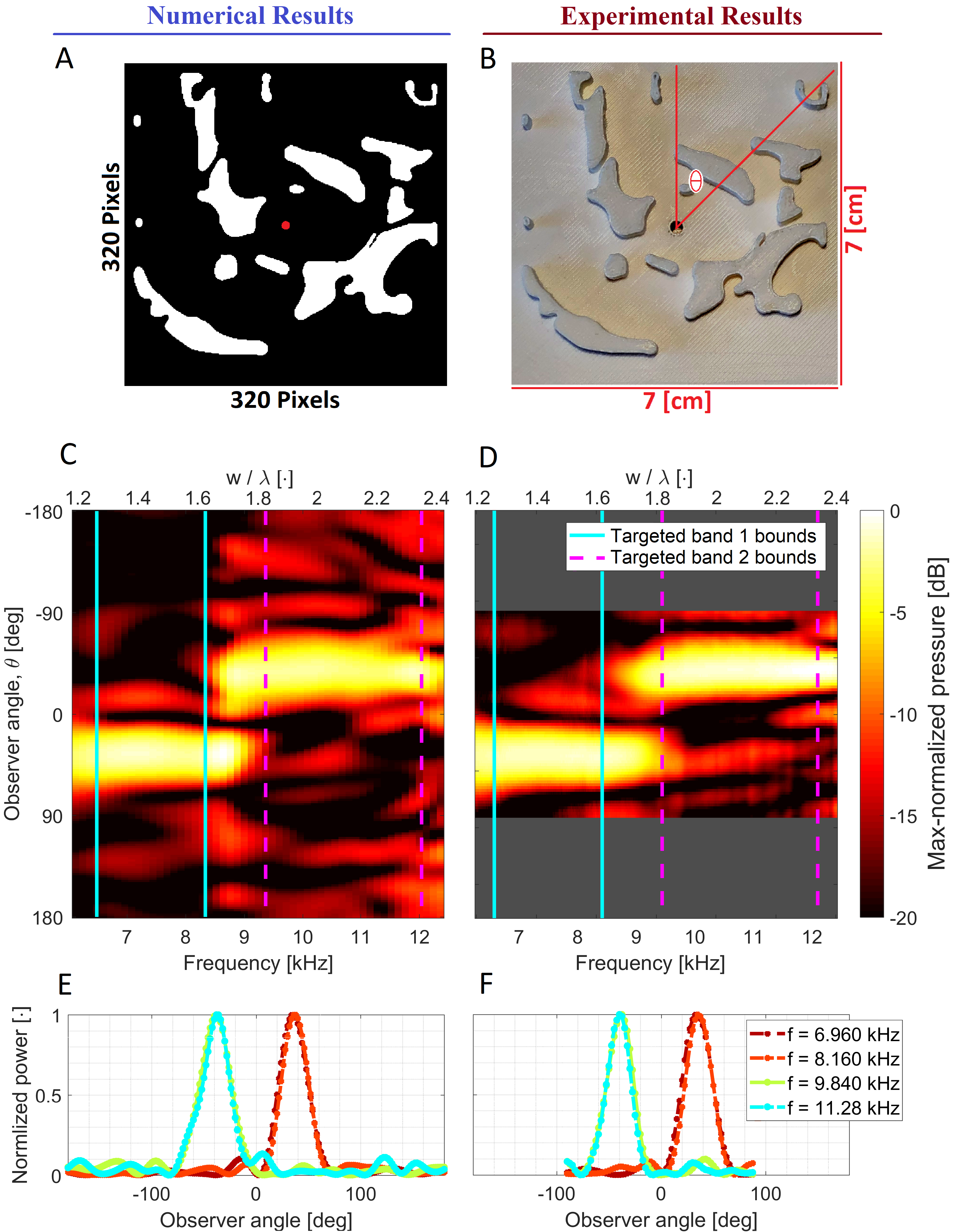}
	\caption{Lambda-splitter geometry and emitted sound pressure: (left column) numerical results and (right column) experimental results. (A) Blueprint for the lambda-splitter designed with the morphogenetic approach, the red dot indicating the source position. (B) Experimentally constructed sample (3D printed) with corresponding dimensions. (C-D) Max-normalized far field pressure maps. The cyan and magenta vertical lines indicates the bounds of the frequency bands targeted in the design process. (C) Simulated sound emission, the majority of the energy of all spectral components are observed to be directed into the +35$^{\text{o}}$ direction or the -35$^{\text{o}}$ direction depending on frequency. (D) Experimental measurements performed on the 3D-printed emitting device showing remarkable agreement with simulations (see Supplementary Information for data treatment procedure). (E-F) Per frequency max-normalized far field power, showing a clear decomposition of the spectral components. (E) Simulated far field emission. (F) Experimental measurements.}
	\label{FIG:ACOUSTIC_WAVESPLITTER_INVESTIGATION}
\end{figure*}

\subsection*{The Acoustic Lambda-splitter}
The spatio-spectral decomposition of sound can be tailored as desired using the proposed design approach. To demonstrate this, we also synthesize a free-space broadband acoustic lambda-splitter. This device separates the sound emitted by a source in two distinct spatial directions, depending on the spectral content of the waves. The lambda-splitter is designed to emit sound in the +35$^{\text{o}}$ direction for frequencies in the band from 6.5 kHz to 8.4 kHz and the -35$^{\text{o}}$ direction for frequencies in the band from 9.4 kHz to 12 kHz. The resulting blueprint for the lambda-splitter and a 3D-printed realization are shown in Fig.~\ref{FIG:ACOUSTIC_WAVESPLITTER_INVESTIGATION}a and Fig.~\ref{FIG:ACOUSTIC_WAVESPLITTER_INVESTIGATION}b, respectively. Maps of the radiated sound pressures are seen in Fig.~\ref{FIG:ACOUSTIC_WAVESPLITTER_INVESTIGATION}c (simulation) and Fig.~\ref{FIG:ACOUSTIC_WAVESPLITTER_INVESTIGATION}d (experiment). The emitted sound is indeed seen to be separated into the two targeted spatial directions according to the frequency content (the frequency-band limits are highlighted using solid and dashed lines). Again a substantial main- to side-lobe difference is observed for the sound pressure inside the targeted bands of operation. Across the low-frequency band more than 88\% of the power emitted over the measured -90$^{\text{o}}$ to 90$^{\text{o}}$ interval is directed into the main lobe at full width half maximum. For the high-frequency band 82\% of the total power is emitted in the target direction. In practice, this is a clearly detectable decomposition of the sound-field. Figure~\ref{FIG:ACOUSTIC_WAVESPLITTER_INVESTIGATION}e (simulation) and Fig.~\ref{FIG:ACOUSTIC_WAVESPLITTER_INVESTIGATION}f (experimental) show the max-normalized far-field power as a function of emission angle for four frequencies; two inside each frequency band. The emission direction of the main-lobe is well separated for frequencies in different bands, and overlaps almost perfectly for frequencies in the same band.

Note that the presented designs are non-unique in the sense that several different design topologies was found to lead to very similar responses in terms of minimizing the goal function. This is indeed expected and often observed when applying optimization techniques to solve inverse design problems as they are (almost) all non-convex in nature. 

\subsection*{Feature Sensitivity}
The complex geometries of the optimized devices, coupled with the wave nature of the physics, means that providing a simple interpretation of the effect of any individal feature on the device performance is functionally impossible. Indeed, one could be tempted to claim that the designs features are simply somewhat randomly distributed and question their importance. 

\begin{figure*}[t]
	\centering
	\includegraphics[width=.7\linewidth]{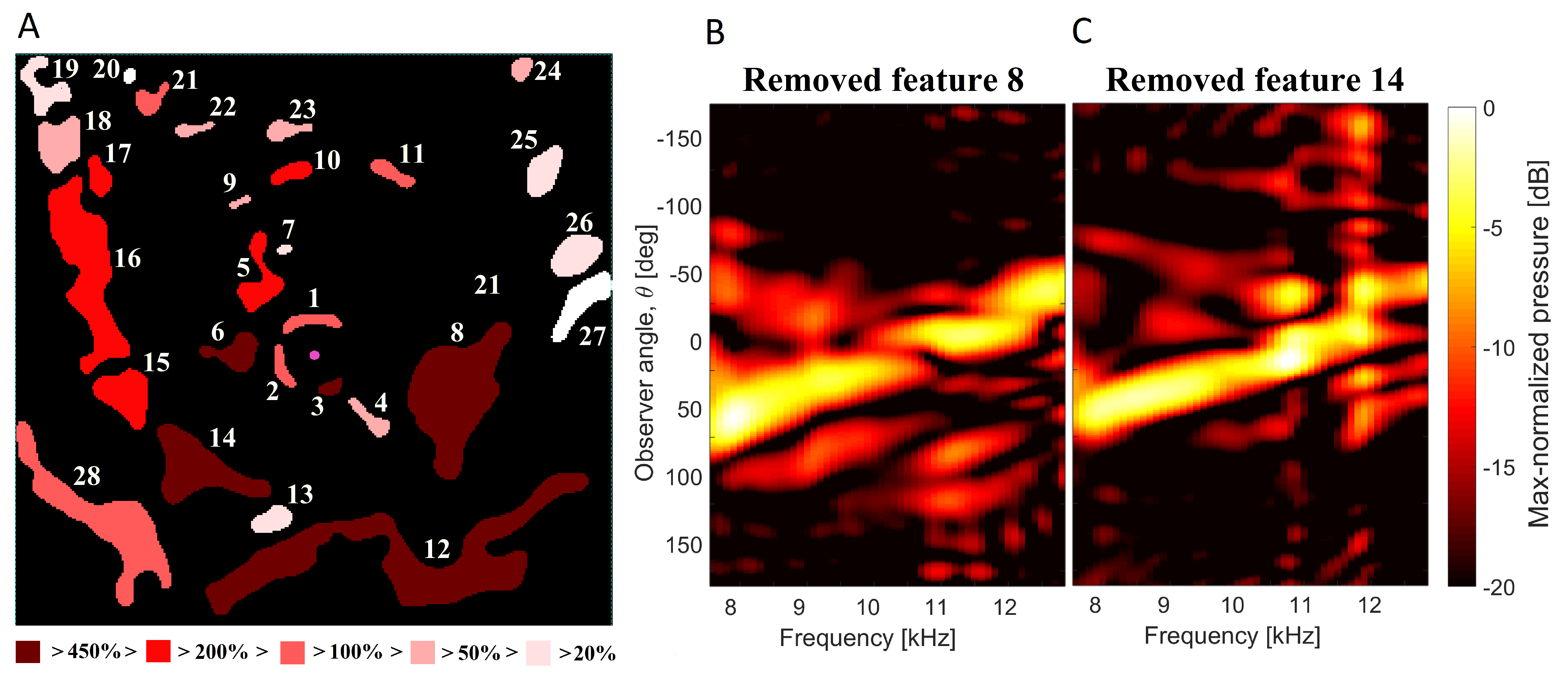}
	\caption{(a) Blueprint for the ARE with numbering of the 28 features and a colormap illustrating the deterioration of $\Phi$ as each single feature is removed showing that (almost) all features have significant importance for $\Phi$. (b-c) Far-field pressure map for the ARE with (b) feature 8 removed and (c) feature 14 removed showing the deterioration of the emission pattern as individual features are removed. (Compare these to the original ARE radiation emission pattern in Fig. \ref{FIG:ACOUSTIC_RAINBOW_INVESTIGATION}c.)\label{FIG:FEATURE_SENSITIVITY_STUDY}}
\end{figure*}

To investigate and illustrate the importance of each of the 28 sub-wavelength features constituting the ARE, we remove the features one-by-one and compute the resulting detrioration of $\Phi$. The results of this investigation are reported in Fig.~\ref{FIG:FEATURE_SENSITIVITY_STUDY}a showing the relative derioration (increase) of $\Phi$ as each feature is removed. The effect is seen to range from a few percent (e.g. removing feature 27) to an effect of more than $450\%$ (e.g. removing feature 6). For reference, the removal of the enitre ARE results in a $\approx 325\%$ deterioration of $\Phi$, while removing the source results in a $\approx 350\%$ deterioration. That is, $\Phi$ is more sensitive to the interplay between certain of the design features than it is to entirely removing the ARE or turning off the sound. From Fig.~\ref{FIG:FEATURE_SENSITIVITY_STUDY}a it is seen that the most important features are generally the largest, the ones situated closest to the source and the features that form the anterior of the ARE. The latter are responsible for the largest phase shifts and the greatest manipulation of the acoustic propagation path, while the smaller features close to the source cover a large part of the solid angle as seen from the source, hereby interacting with the largest fraction of the emitted power. In contrast, removing the smaller features sitated furthest from the source above and to the left are generally seen to have the smallest effect on $\Phi$. This, likely due to the fact that they are exposed to the smallest amount of total emitted acoustic power thus having less of an effect on overall field. The effect of removing particular features on the emission pattern are shown in Fig.~\ref{FIG:FEATURE_SENSITIVITY_STUDY}b (feature 8) and Fig.~\ref{FIG:FEATURE_SENSITIVITY_STUDY}c (feature 14). Comparing these maps to the one for the optimized device in Fig.~\ref{FIG:ACOUSTIC_RAINBOW_INVESTIGATION}c it is seen that the changes to the radiation pattern are quite complex. Removing feature 8 is seen to affect the emission pattern significantly across the majority of the frequency band of operation, while removing feature 14 mainly affects the pattern at frequencies above 10 kHz.

\section*{Conclussion}

The design and experimental realization of a device forming an acoustic rainbow in free space reveal an uptapped potential for manipulating the spatio-spatial properties of sound fields based solely on passive scattering elements. The utilization of passive spatio-spectral sound manipulation with high emission efficiency opens an avenue for reducing losses currently associated with active sound manipulation. In addition, this work demonstrates a powerful approach that could be used to aid in the understanding of the acoustic radiation and scattering processes found in complex natural structures. The principles explored in this work are of relevance for acoustic transduction, sensor design, bionics, optics and other diverse applications, such as radio backscattering. 

Our two sample applications of controlled morphogenesis of sound are quite diffrent in terms of the performance goal, yet both scattering structures show distinct angular geometric asymmetries. These asymmetries account for the changing length of the sound propagation paths as a function of direction, leading to the interaction phenomena required for the spatio-spectral decomposition of sound in the sub-wavelength range. This leads to the observation that some of these structural features are reminiscent of features commonly found in human pinnae, in particular the acoustic channel formed by the helix and the anti-helix, as well as the tragus and anti-tragus of smaller size than other features, and in proximity to the entrance of the ear canal. The dimensions of the pinnae features and the frequencies they act upon ( $> 6$ kHz) match those of the scattering devices presented here, suggesting a potential relation between some of the evolutionary features that aid human sound source localization, and the morphogenesis of sound realised in this work. 

\section*{Materials and Methods} 
The design method was implemented using an in-house MATLAB code, where the physics was modelled using an implementation of the Helmholtz equation \cite{BOOK_FUNDAMENTAL_ACOUSTICS}. The frequency response of the optimized structures was evaluated with COMSOL Multiphysics \cite{COMSOL61main} using both  2D and full 3D models. The ARE and lambda-splitter were fabricated in hard plastic using 3D-printing. Details about the proposed morphological design method, the numerical model, the experimental setup and experimental procedure, the procedure for creating Fig.~\ref{FIG:ACOUSTIC_RAINBOW_ILLUSTRATION}, as well as a set of video and audio examples of the sound emission from the two devices are available in the Supplementary Information.

\section*{Data availability}
The raw experimental data, experimental setup specifications, datasheets for the materials used for the 3D printed devices, blueprints for the devices as well as detailed information about the material, model and optimization parameters used in the design process are available from the corresponding author upon reasonable request.

\section*{Code availability}
The computational models used to evaluate the devices as well as the code used in the post processing of data are available from the corresponding author upon reasonable request.

\section*{Acknowledgements}
Senior Researcher Fengwen Wang and Technician Jannick Schultz performed 3D printing of the ARE and lambda-splitter at the Technical University of Denmark. Assistant Engineer Jørgen Rasmussen aided in the preparation of the experimental setup. J.J. Thomsen from the Technical University of Denmark contributed with input and discussions during the manuscript development. VILLUM Fonden funded this work via the NATEC (NAnophotonics for Terabit Communications) Centre of Excellence (grant no. 8692) and the INNOTop (InnoTop - Interactive, Non-Linear, High-Resolution Topology Optimization) Villum Investigator project.

\section*{Author Contributions}
R.E.C. contributed to the original idea, method development, software implementation, design process, experimental setup design, experimental measurements, data processing, analysis of results and manuscript preparation. O.S. contributed to the original idea, method development, analysis of results, manuscript preparation and funding acquisition. E.F. contributed to the original idea, experimental setup design, analysis of results and manuscript preparation.

\section*{Competing Interests}
The authors declare no competing interests.

\section*{Additional Information}
Supplementary information is available for this paper in the Ancillary files.

\section*{Materials  \&  Correspondence}
Correspondence and requests for materials should be addressed to R.E.C.

\subsection*{References}

\bibliography{References}

\end{document}